\documentclass[doublecol]{epl2} 

\newcommand{\be}{\begin{equation}}
\newcommand{\ee}{\end{equation}}
\usepackage{amsmath}

\title{Herschel-Bulkley rheology from lattice kinetic theory of soft-glassy materials}
\shorttitle{Herschel-Bulkley rheology from lattice kinetic theory} 

\author{R. BENZI\inst{1}, M. BERNASCHI\inst{2}, M. SBRAGAGLIA\inst{1}, AND S. SUCCI\inst{2}}
\shortauthor{R. Benzi \etal}

\institute{                    
  \inst{1} Department of Physics and INFN, University of Tor Vergata, Via della Ricerca Scientifica 1, 00133 Rome, Italy \\
  \inst{2} Istituto per le Applicazioni del Calcolo CNR, Via dei Taurini 9, 00185 Roma, Italy.}


\abstract{We provide a clear evidence that a two species mesoscopic Lattice Boltzmann (LB) model with competing short-range attractive and mid-range repulsive interactions supports emergent Herschel-Bulkley (HB) rheology, i.e. a power-law dependence of the shear-stress as a function of the strain rate, beyond a given yield-stress threshold. This kinetic formulation supports a seamless transition from  flowing to non-flowing behaviour, through a smooth tuning of the parameters governing the mesoscopic interactions between the two species. The present model may become a valuable computational tool for the investigation of the rheology of soft-glassy materials on scales of experimental interest.}

\begin{document}

\maketitle

The rheology of soft flowing systems, such as emulsions, foams, pastes gels, and other types of complex fluids, plays a major role in modern materials science, both on account of its broad range of practical applications and because of the  challenges it poses to modern non-equilibrium thermodynamics \cite{COMPLEX1,COMPLEX2,COMPLEX3,COMPLEX4}. Soft-glassy materials of assorted nature, emulsions, foams, pastes and granular materials, are known to exhibit a fairly rich and complex rheology. Among other signatures of complex behaviour, such as anomalous relaxation, dynamical arrest and refluidization, stick and slip motion, the rheology of soft-glassy materials is often characterized by a non-linear relation between the applied stress and the resulting strain. A popular expression of such non-linear behaviour is provided by the Herschel-Bulkley (HB) relation \cite{HB1,HB2,HB3,PASTE}, $\sigma = \sigma_Y + A S^{\beta}$, where $\sigma$ is the applied stress, $S$ the resulting shear (inverse time) and $A$ a material constant. The HB relation is characterized by a non-zero yield-stress, $\sigma_Y$, below which no flow takes place, and by a scaling exponent $\beta \ne 1$. Although non-linear rheological behaviour is well documented in several experimental studies, its microscopic foundations still elude a thorough theoretical understanding, thereby holding back many important applications in fluid mechanics, material science and biology. As for most complex states of matter, the experimental and theoretical investigation of soft-glassy materials draws substantial benefits from the additional insights provided by computer simulations.  Simulation methods split into two major families: macroscopic/continuum and microscopic/atomistic. The former are computationally efficient, but require a certain fore-knowledge of the basic physics in order to supply, upfront, constitutive equations and boundary conditions. Microscopic methods require much less coarse-graining, and, as a consequence, less parametric input, but must face with a much higher computational demand. A third option is offered by mesoscopic methods, which, as implied by their very name, work at an intermediate level, hopefully achieving an optimal tradeoff between the aforementioned two. Mesoscopic models supporting HB rheology have been in existence for a while in the soft glassy materials literature, to begin with the well-known model by Sollich {\it et al.}, in which the authors postulate a model kinetic equation for the probability $P(l,E,t)$ of finding a given mesoscopic region of the flow at time $t$, with a local strain $l$ and a local maximal yield elasticity $E$ \cite{SOLLICH97,SOLLICH98,FIELDING}. More recently, kinetic models for the elastoplastic dynamics of jammed materials, taking the form of non-local Boltzmann equations for the stress distribution function have also been proposed \cite{BOCQUET}. 
In this work, we provide the first evidence that a mesoscopic Lattice Boltzmann (LB) model with competing short-range attractive and mid-range repulsive interactions supports emergent Herschel-Bulkley (HB) rheology, i.e. a power-law dependence of the shear-stress as a function of the strain rate, beyond a given yield-stress threshold. The kinetic equation describing the fluid rheology is not postulated on the basis of informed insights on the physics under inspection, but results instead from a lattice transcription of a basic Boltzmann  kinetic equation, equipped with some minimal ingredients required to reproduce the hydrodynamics of non-ideal fluid mixtures  \cite{PRLnoi,JCPnoi}.  

\begin{figure}
\begin{center}
\includegraphics[scale=0.3]{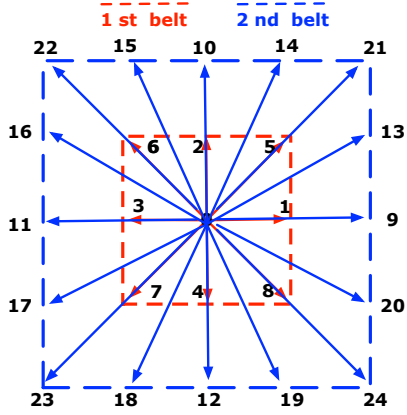}
\end{center}
\caption{The set of $25$ discrete velocities, including a rest particle ($0$). The first belt ($1-8$) hosts attractive AA and BB and repulsive AB interactions. Both belts ($1-24$) host repulsive AA and BB interactions. Full details coan be found in \cite{JCPnoi}.}
\label{FIG4}
\end{figure}

\section{Lattice Boltzmann with multirange interactions}

Our system is described by a lattice version of the Boltzmann kinetic equation for a multicomponent fluid \cite{Benzi92,SC_93,SD_95,BGK54,Gladrow00} with two species ($A,B$):
\begin{eqnarray}
f_{is}(\vec{r} + \vec{c}_i, t + 1)-f_{is}(\vec{r} , t ) =\hspace{.2in} \nonumber  \\
-\frac{1}{\tau}[f_{is}(\vec{r} , t )-f_{is}^{(eq)}(\rho_s ,  \vec{u}+\tau \vec{F}_{s} /\rho_s)],   \hspace{.2in} s=A,B
\label{eq:be}
\end{eqnarray}
where $f_{is}(\vec{r},t)$ is the probability density function of finding a particle of species $s=A,B$ at site $\vec{r}$ and time $t$, moving along the $i$-th lattice direction defined by the discrete speeds $\vec{c}_i$ with $i=0, 1,..., b=24$ (see figure  \ref{FIG4}). The left hand-side of (\ref{eq:be}) stands for molecular free-streaming, whereas the right-hand side represents the collisional relaxation towards a local equilibrium $f_{is}^{(eq)}(\rho_s , \vec{u} )$ on a time scale $\tau$. The equilibrium for the $s$ specie is a function of the local species density (one for each species) and of the baricentric velocity $\vec{u}$: 
$$\rho_s(\vec{r}, t)=\sum_i f_{is}(\vec{r}, t); \hspace{.2in} \vec{u} (\vec{r}, t)= \frac{\sum_s \sum_i f_{is}(\vec{r}, t) \vec{c}_i } { \sum_s \rho_s (\vec{r}, t)}  $$
$$
f_{is}^{(eq)}(\rho_s,\vec{u} )=w_i \rho_s \left(1+\frac{\vec{u} \cdot \vec{c}_{i}}{c_S^2}+ 
\frac{\vec{u} \vec{u} : {\vec{c}_{i} \vec{c}_{i}-c_S^2 I }}{2 c_S^4}\right)
$$
where $c_S^2=\sum_i w_i c_{ix}^2$ is the square of the sound speed velocity, 
$I$ is the unit tensor and $w_i$'s are equilibrium weights used to enforce isotropy of the hydrodynamic equations \cite{Benzi92}. Intermolecular forces are incorporated within the shift $\vec{F}_s \tau/\rho_s$ in the baricentric velocity in (\ref{eq:be}). 
The force within each species, $\vec{F}_s$, consists of an attractive ($a$) component , acting only on the first Brillouin region ($b_1$,  index $1-8$ in figure \ref{FIG4}), and a repulsive ($r$) one, acting on both belts ($b_2$, index $1-24$ in figure \ref{FIG4}), whereas  the force between different species ($X$) is short-ranged and repulsive (acting again on the first Brillouin region):  
\be\label{FFF}
\vec{F}_s(\vec{r}, t) = \vec{F}^a_s(\vec{r}, t) + \vec{F}^r_s(\vec{r}, t)+\vec{F}^X_s(\vec{r},t)
\ee
with the general structure of the forcings given by
$$
\vec{F}^{a,r}_{s}(\vec{r},r) = -G^{a,r}_{s} \psi_{s}({\vec r},t) 
\sum_{i \in b_{1,2}} w_{i} \vec{c}_i \psi_{s}({\vec r}+{\vec c}_i,t) 
$$
$$
\vec{F}^{X}_{s}(\vec{r},r) = -\frac{G^{X}_{ss^{\prime}}}{\rho_0} \rho_{s}({\vec r},t) 
\sum_{i \in b_1} w_{i} \vec{c}_i \rho_{s^{\prime}}({\vec r}+{\vec c}_i,t), \hspace{.2in}    s \neq s^{\prime} 
$$
with $w_i$ the standard weights of the two-dimensional nine-speed lattice, $G$'s the strength parameters. The pseudo-potential  $\psi_{s}({\vec r},t)$ l has been taken for both species in the form originally 
suggested by Shan \& Chen \cite{SC_93}, namely 
$
\Psi(\rho_s) = \rho_0 (1-e^{-\rho_s/\rho_{0}})
$. 
The parameter $\rho_0$ is a reference density beyond which self-interactions 
become vanishingly small, thereby preventing mass density collapse 
(i.e. $\rho_s \rightarrow \infty$) due to attractive interactions. 
Two-belt, (intra-species) self-interactions are introduced to allow a separate control 
of the equation of state and surface tension, independently. 
In particular, one can show that, for a flat A/B interface, the surface tension scales like:
$$
\gamma \propto - \sum_s \tilde{G}_s  \int |\nabla \psi_s|^2 dy - 
\frac{G_{AB}}{\rho_0^2} \int \nabla \rho_A \cdot \nabla \rho_B \; dy.
$$
where the coordinate $y$ runs across the interface and $\tilde{G}_s \equiv G^a_{ss}+ {12\over 7} G^r_{ss}$.
For repulsive interactions, ($G_{AB} >0$), the second integral at the rhs is 
positive-definite, since $\nabla \rho_A \cdot \nabla \rho_B <0$. 
By choosing $\tilde{G}_s >0$, the first integral is negative-definite and 
consequently one can decrease the surface tension by simply increasing $\rho_0$.  
Full details can be found in \cite{JCPnoi}.  As is well known, non-trivial rheological behaviour has been obtained by molecular dynamics simulation models \cite{MDMOD,MDMODb}. A basic lesson learned from these models is that by taking two fluids with suitable interaction parameters (involving frustration), one is able to observe a phenomenology in  reasonable qualitative agreement with experimental results. This suggests the possibility of formulating an equivalent model at the level of a suitably extended kinetic Boltzmann equation with minimal ingredients (two species plus frustration) to support non-linear rheology.  This is exactly what characterizes our model. The present LB scheme embeds the {\it universality} of the conservation laws underlying the fluid 
equations, be they ideal or interacting (non-ideal), within a computationally
efficient theoretical framework.
We note that the Shan-Chen formulation is basically an effective 
one-body closure of the many-body Liouville equation, encoding the basic symmetries 
of potential energy interactions within a minimal lattice
formulation, i.e. a one-parameter, nearest-neighbor, pseudo-potential.   
The reason why our model can incorporate substantial new non-ideal physics 
without taxing computational efficiency, is again universality: once the proper 
competing mechanisms are put in place, the specific form of the interactions is 
largely immaterial to the large-scale behaviour of the non-ideal fluid. 
Consequently, a minimal lattice pseudo-potential is sufficient. 
We remark that a unique feature of the present LB scheme, is the capability of incorporating 
non-linear hydrodynamics nearly "for-free", through a simple quadratic dependence of the 
local equilibria on the local flow field. 
Thanks to this property, our model can seamlessy straddle across various 
non-trivial flow regimes (flowing/arrested) through a smooth change of the interaction parameters.

\section{Numerical Results}

The computational domain is a square box of size $L \times L$ covered by $N_x \times N_y =512 \times 512$ lattice sites with a uniform lattice spacing $dx=1$. The simulations, performed on latest generation Graphics Processing Units (GPU) \cite{Bernaschi09}, require few hours for one million time-steps, the typical time-span of a run.  With a fixed set of following baseline coupling parameters  \footnote{We have chosen $G^a_{AA}=-9.,G^r_{AA}=8.1, G^a_{BB}=-8,G^r_{BB}=7.1, G^r_{AB}=0.045$. Negative/positive signs standing for repulsion/attraction, respectively, secure that both $A$ and $B$ fluids are in the liquid phase. For all simulations we have chosen a constant relaxation time $\tau=1.0$. 
The use of a coupling-dependent relaxation time has never been explored in the literature 
and surely deserves a separate study on its own.}, the reference density $\rho_0$  is varied between $\rho_0=0.70$ and $0.90$ that corresponds to a decrease of surface tension from ordinary values to an almost vanishing value for $\rho_0=0.90$ (based on the use of equation (63) in \cite{JCPnoi}).  The  fluid is initialized with $\rho_A(x,y)=0.61 (1+0.1 \, \sin (k_{in} y))$ and $\rho_B(x,y)=0.61 (1+0.01 \, \sin (k_{in} y))$ with $k_{in}=\frac{64 \pi}{L}$, and is subject to an external periodic forcing in the $x$ direction of the form $F_x(y) = F_0 \sin (k_f y)$, with wavenumber $k_f=\frac{2 \pi}{L}$. The forcing amplitude $F_0$ is tuned  in such a way as to produce, in standard stationary flow conditions, a sinusoidal Kolmogorov flow of maximum speed $U_0$, i.e. $u_x(x,y)=U_0 \sin(k_f y)$.  In a previous work \cite{JCPnoi}, the system response was monitored using the following response function: 
$R(t)=\frac{\hat{\bar U}(t)}{U_0} \equiv \frac{\nu_0}{\bar{\nu}}$ where $\hat{\bar U}(t)$ is 
the Fourier transform of the line-averaged speed along the $x$ direction
$$U_x(y,t)=\frac{\sum_x u_x(x,y,t)}{L}$$ 
$$\hat{\bar U}(t)= \frac{2 \sum_y U_x(y,t) \sin(k_f y)}{L}.  $$

\begin{figure}[h]
\begin{center}
\includegraphics[width=0.40\textwidth]{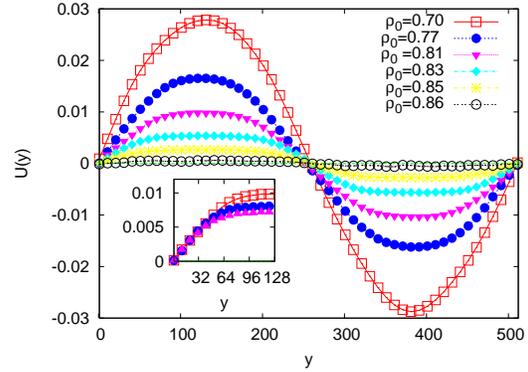}
\end{center}
\caption{The average (time and $x$ direction) velocity profile $U(y)$ for different values of $\rho_0$: by increasing $\rho_0$ (i.e. decreasing the surface tension) the velocity profile becomes flatter and with a lower amplitude. In the inset, we show $U(y)$ for $\rho_0= 0.79$, $\rho_0=0.81$ and $\rho_0=0.83$ rescaled in such a way that the velocity gradient at $y=0$ is kept fixed. The inset shows that the different velocity profiles cannot be superimposed by a mere rescaling factor.
}
\label{FIG1}
\end{figure}

In the above,  $\nu$  is the nominal kinematic viscosity of both fluids and $\bar{\nu}$ defines the effective  viscosity of the two-fluids system.  By construction, under undisturbed flow conditions,  $R=1$, so that $R\ll 1$ provides a  direct measure of slowing-down through enhanced effective viscosity. The parameter $R$ is thus a direct measure of the effective fluidity of the system. However, since we are focusing on a non-Newtonian behaviour, it proves more informative to inspect first the actual space-time averaged velocity profiles 
$${U}(y)=\frac{1}{T}\int_0^T U_x(y,t) dt,  \hspace{.2in}  T \gg 1$$ 
as a function of the reference density $\rho_0$ at a given forcing intensity with $U_0=0.1$ in computational units. From figure  \ref{FIG1}, a flattening of the velocity profile in the central region of the flow is clearly observed, for all values of $\rho_0>0.70$.  This is a well-known signature of non-Newtonian behaviour \cite{GOYON}. 

\begin{figure}[h]
\begin{center}
\includegraphics[width=0.40\textwidth]{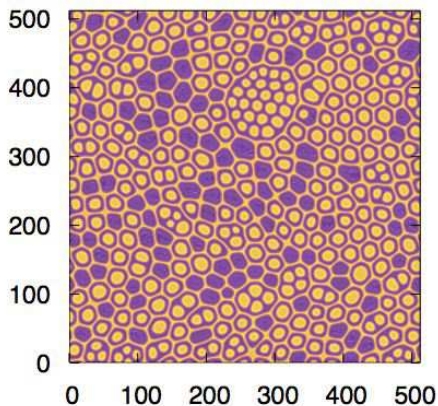}
\end{center}
\caption{A typical snapshot of the fluid $A$ density contour. Blue/yellow colors code for low/high density regions. Low/high density regions of fluid $A$ are filled with correspondingly high/low density regions of fluid $B$.
}
\label{FIGBOLLE}
\end{figure}

\begin{figure}
\begin{center}
\includegraphics[width=0.27\textwidth,angle=-90]{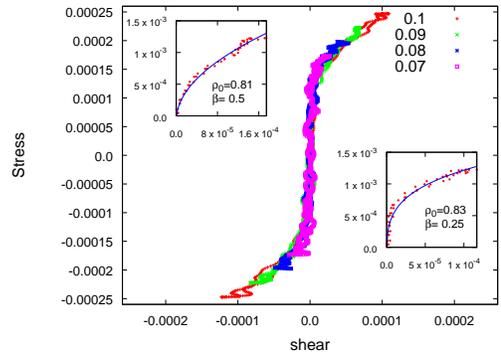}
\end{center}
\caption{
The average stress $P_{xy}(y)$ as a function of the observed shear
$S(y)$ obtained from the numerical simulation at $\rho_0=0.83$ and
different values of the forcing.
In the inset, we show the fit of $P_{xy}$ by using the HB form
$A+B \;S^{\beta}$ for $\rho_0=0.83$ and $\rho_0=0.81$.
}
\label{FIG2}
\end{figure}

A typical density contour of fluid $A$ is shown in figure \ref{FIGBOLLE}. To inspect the non-Newtonian behaviour on more quantitative grounds, we have measured the effective viscosity through the ratio of the nominal shear for a standard flow, to the value of the shear $S(y) \equiv d U/dy$ provided by the simulation (see inset of figure \ref{FIG1}).  At statistical steady state, the momentum balance equation yields $\partial_y P_{xy} = F_x(y)$ (derivatives along $x$ are zero by homogeneity). Integrating along $y$, we obtain $P_{xy}(y) = \int_{0}^y F_x(y') dy'$, which is known exactly at each location $y$, since the right-hand-side is nothing but the expression of the forcing. The resulting shear is simply collected as the spatial derivative of the time averaged velocity field, i.e. $S_{xy}(y) = \partial_y U(y)$. Figure \ref{FIG2} shows the scatter-plot of the stress $\sigma \equiv P_{xy}$ versus the shear $S \equiv S_{xy}$ for each value of $y$. This figure carries the central result of this work. First, it is seen that the fluid starts to flow only above a critical threshold (yield-stress) of the order of $\sigma_Y \sim  1.5 \; 10^{-4}$, which is comparable with the maximum applied stress $ \sigma_0 \equiv \rho \nu U_0 2 \pi/L \sim 2.5 \times 10^{-4}$. Remarkably, the various data, corresponding to different values of the forcing, all fall within basically the same master curve. In the lower inset, we report the fit exposing the exponent of the HB-like relation $\sigma= \sigma_Y + B \; S^{\beta}$, which yields $\beta \sim 0.25$, in a reasonable good agreement with previous models \cite{PREVIOUS,PREVIOUSb}. The upper inset shows the same fit for $\rho_0=0.81$, which again yields HB behaviour, although with a larger exponent $\beta \sim 0.5$.

\begin{figure}[h]
\begin{center}
\includegraphics[width=0.40\textwidth]{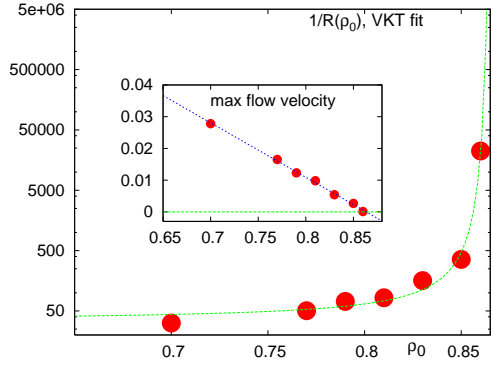}
\end{center}
\caption{
The average ``response'' $R(\rho_0)$ for different $\rho_0$. Assuming that $ R \sim \frac{1}{\bar{\nu}}$, where $\bar{\nu}$ is the effective viscosity, we plot the reciprocal response $1/R$ versus $\rho_0$. The continuous line is the best fit using $R^{-1}(\rho_0) = A\,  exp \left(\frac{B}{\rho_c-\rho} \right)$, with $\rho_c = 0.867$. In the inset we show the value of the maximum mean velocity.}
\label{FIG3}
\end{figure}

Since our data support HB behaviour with a surface-tension dependent exponent, it is worth inspecting the effect of lowering the surface tension, and eventually taking it nominally below zero. To this purpose, we measure the time-averaged response function $R$ for different values of $\rho_0$. Figure \ref{FIG3} shows a neat divergence of the reciprocal response function as the condition of zero-flow (total arrest) is approached. Incidentally, the functional dependence of the  time averaged response function, $R(\rho_0)$, can be fitted reasonably well by a Vogel-Fulcher-Tammann (VFT) law \cite{VFT1,VFT2,VFT3}, $R^{-1}(\rho_0) = A \; exp(\frac{B}{\rho_c-\rho})$, with $A=3.51$ and $B=0.045$, although other functional forms compatible with finite-density divergence cannot be ruled out. For instance, the value of the maximum mean velocity shown in the inset of figure \ref{FIG3}, would support a simpler $R^{-1} \sim (\rho-\rho_c)^{-1}$ divergence. Leaving this question to a future and separate investigation, here we simply observe that the system appears to come to a complete arrest as the surface tension is sent to smaller and smaller values (the nominal zero-point is at $\rho_0 \sim 0.87$). Finally, we point out that the system can also be taken to virtually negative surface tensions, in which case lamellar-like configurations are observed. However, the physical viability/reliability of the present model in this parameter regime still needs to be assessed.

\section{Conclusions and Outlook}

Summarizing, we have provided the first evidence of emergent Herschel-Bulkley (HB) rheology from a \lq \lq first principle" lattice kinetic model incorporating the basic ingredients of non-ideal fluids with competing attractive/repulsive interactions.  Although a one-to-one mapping with a corresponding physical system remains to be developed, the present model exhibits a number of highly non-trivial features of soft-glassy behaviour, including the Herschel-Bulkley rheology discussed in this Letter. Finally, in  light of the results discussed in this paper, one could raise the following questions: how far are present materials/experiments  from the scenario depicted in this Letter?   Can new materials/conditions be adapted/designed in such a way as   to realize the scenario revealed/suggested by the simulations?    Since the present mesoscopic model can access scales close to experimental ones,  we hope  that the present work can raise new  stimulating challenges for joint numerical/experimental work.

\acknowledgments

Valuable discussions with H.C. Oettinger, H.J. Herrmann and I.V. Karlin are kindly acknowledged.

\acknowledgments

\end{document}